\documentclass[12pt]{article}

\title{Hunting for the alpha: $B\to \rho\rho$, $B \to \pi\pi$, $B
\to\pi\rho$}
\author {G.G. Ovanesyan\thanks{ovanesyn@itep.ru}, \\ {\it Moscow Institute of
Physics and Technologies and ITEP, Russia} \\ M.I.
Vysotsky\thanks{vysotsky@itep.ru} \\ {\it ITEP, Russia}}
\date{}

\begin{document}
\maketitle

\begin{abstract}

The hypothesis of the smallness of penguin contribution to
charmless strangeless $B_d (\bar B_d)$ decays allows to determine
with high accuracy the value of angle $\alpha$ from the currently
available $B \to \rho\rho$, $B \to \pi\pi$ and $B\to \rho\pi$
decay data.
\end{abstract}

\section{Introduction}

Measurement of CP asymmetries in $B_d(\bar B_d) \to J/\psi K^0$
decays by BaBar and Belle collaborations determines angle $\beta$
of CKM unitarity triangle with high accuracy \cite{1}:
\begin{equation}
\sin 2\beta = 0.724 \pm 0.040 \; , \;\; \beta = 23^o \pm 2^o \;\;
. \label{1}
\end{equation}

The next task is to measure the angles $\alpha$ and $\gamma$ with
comparable accuracy in order to determine if New Physics
contribute to CP-violation in $B$ decays. For precise
determination of the value of angle $\gamma$ one should study
$B_s$ decays and this should wait until LHC(b) era. The purpose of
this Letter is to stress that (may be) angle $\alpha$ is already
known with the accuracy comparable to that achieved in $\beta$.

\section{$B \to \rho\rho$}

Let us start from $B_d(\bar B_d) \to \rho\rho$ decays, where the
smallness of QCD penguin contribution directly follows from
experimental data on relative smallness of the branching ratio of
$B_d(\bar B_d) \to \rho^0\rho^0$ decays \cite{2}. Here are the
experimental data; all the branchings are in units of $10^{-6}$:
\begin{equation}
\begin{array}{lllll} Br(\rho^+ \rho^-) \equiv B_{+-} & = & 30 \pm 5 \pm 4 & \mbox{\rm
\cite{3}} &
\\
Br(\rho^\pm \rho^0) \equiv B_{\pm 0}& = & 22.5 \pm 5 \pm 6 &
\mbox{\rm \cite{4}} & Br(\rho^0 \rho^0) \equiv B_{00} < 1.1
(\rm{90\% ~ C.L.}) \;\; \mbox{\rm \cite{6}} \\ & & 31.7 \pm 7 \pm
5 & \mbox{\rm \cite{5}} &
\end{array}
\label{2}
\end{equation}

In order to prove the smallness of the penguin contribution let us
write the amplitudes of $B_d(\bar B_d) \to \rho^0\rho^0$ decays as
the sum of tree and penguin contributions:
\begin{eqnarray}
A_{\rho^0 \rho^0} & = & T_{\rho^0 \rho^0} e^{i\gamma} + P_{\rho^0
\rho^0} e^{i\delta_{00}} \;\; , \nonumber \\ \bar A_{\rho^0
\rho^0} & = & T_{\rho^0 \rho^0} e^{-i\gamma} + P_{\rho^0 \rho^0}
e^{i\delta_{00}} \;\; , \label{3}
\end{eqnarray}
where $\gamma$ is the angle of a unitarity triangle and
$\delta_{00}$ is the difference of phases of the final state
strong interaction amplitudes induced by the penguin and tree
quark diagrams. (We use the so-called $c$-convention in defining
penguin amplitude: penguin with an intermediate $t$-quark is
subtracted while penguin with an intermediate $u$-quark is
included into the tree amplitude.)

For widths we get:
\begin{eqnarray}
\Gamma_{\rho^0 \rho^0} & = & T_{\rho^0 \rho^0}^2 + P_{\rho^0
\rho^0}^2 + 2T_{\rho^0 \rho^0} P_{\rho^0 \rho^0} \cos(\delta_{00}
-\gamma) \;\; , \nonumber \\ \bar\Gamma_{\rho^0 \rho^0} & = &
T_{\rho^0 \rho^0}^2 + P_{\rho^0 \rho^0}^2 + 2T_{\rho^0 \rho^0}
P_{\rho^0 \rho^0} \cos(\delta_{00} +\gamma) \;\; , \label{4}
\end{eqnarray}
and
\begin{equation}
\frac{1}{2}(\Gamma_{\rho^0 \rho^0} + \bar\Gamma_{\rho^0 \rho^0}) =
T_{\rho^0 \rho^0}^2 + P_{\rho^0 \rho^0}^2 + 2T_{\rho^0 \rho^0}
P_{\rho^0 \rho^0} \cos\gamma \cos\delta_{00} \geq P_{\rho^0 \rho^0}^2
(1- \cos^2 \gamma) \;\; . \label{5}
\end{equation}

Since from the global fit of CKM matrix parameters we know that
$\gamma \geq 45^o$ \cite{13}-\cite{15}, one observes that the
compensation of $P_{\rho^0 \rho^0}$ by $T_{\rho^0 \rho^0}$ is not
possible and both of them are small in comparison with the
amplitudes of $B$ decays into $\rho^\pm \rho^0$, $\rho^+
\rho^-$-states.

Two $\rho$-mesons produced in $B$ decays should be in $I = 0$ or
$I = 2$ states, and since QCD penguin amplitude has $\Delta I =
1/2$, it contributes only to $I = 0$ state. That is why
$P_{\rho^\pm \rho^0} = 0$, while $P_{\rho^+ \rho^-} = P_{\rho^0
\rho^0}/\sqrt{2} \ll T_{\rho^+ \rho^-}$. Tree level $b \to u \bar
u d$ amplitude having both $\Delta I = 1/2$ and $\Delta I = 3/2$
parts produce both $I=0$ and $I=2$ states of two $\rho$-mesons,
and one can easily organize compensation of these two amplitudes
in $B_d(\bar B_d) \to \rho^0\rho^0$ decays still satisfactorily
describing $B_d(\bar B_d) \to \rho^+\rho^-$ and $B_u(\bar B_u) \to
\rho^\pm\rho^0$ branching ratios.

Let us show how it works:
\begin{equation}
T_{\rho^0 \rho^0} = \frac{1}{\sqrt{6}}A_0 - \frac{1}{\sqrt{3}} A_2
e^{i\delta} \;\; , \label{6}
\end{equation}
where $\delta$ is the difference of the phases of final state
interaction (FSI) amplitudes of $\rho$-mesons in $I=2$ and $I=0$
states and in order for these two terms to compensate each other
$\delta$ should be small. Let us suppose that $\delta =0$, so that
we can write:
\begin{equation}
\frac{1}{\sqrt{6}}A_0 = \frac{1}{\sqrt{3}}A_2 \pm \sqrt{B_{00}}
\;\; . \label{7}
\end{equation}
We should extract the value of $A_2$ from $B_u(\bar B_u) \to
\rho^\pm\rho^0$ decay branching ratio:
\begin{equation}
T_{\rho^\pm \rho^0} = \frac{\sqrt{3}}{2}A_2 \; , \;\; A_2 =
\frac{2}{\sqrt{3}}\sqrt{B_{\pm 0}k} \;\; , \label{8}
\end{equation}
where $k= \tau_{B^0}/\tau_{B^+} = 0.92$.

Finally we get:
\begin{equation}
T_{\rho^+ \rho^-} = \frac{1}{\sqrt{3}}A_0 +\frac{1}{\sqrt{6}}A_2 =
\sqrt{\frac{2}{3}}A_2 \pm \sqrt{2B_{00}} + \frac{1}{\sqrt{6}}A_2 =
\sqrt{2B_{\pm 0} k} \pm \sqrt{2B_{00}} \;\; , \label{9}
\end{equation}
and choosing a negative sign and the upper experimental bound on
$B_{00}$ as well as an average experimental result for $B_{\pm 0}$
we obtain for $Br(\rho^+ \rho^-)$ the result which coincides with
the central value from (\ref{2}).

Turning to our main subject -- determination of the value of the
angle alpha -- we should look at CP-asymmetries, measured in
$B_d(\bar B_d) \to \rho^+\rho^-$ decays\footnote{This simple
formula is valid only for the decays to longitudinally polarized
$\rho$-mesons; fortunately $f_L = 0.99 \pm 0.03 \pm 0.03$.}:
\begin{equation}
\frac{\frac{dN(\bar B_d^0 \to \rho^+ \rho^-)}{dt} - \frac{dN(B_d^0
\to \rho^+ \rho^-)}{dt}}{\frac{dN(\bar B_d^0 \to \rho^+
\rho^-)}{dt} + \frac{dN(B_d^0 \to \rho^+ \rho^-)}{dt}} =
-C_{\rho\rho}\cos(\Delta m \Delta t) + S_{\rho\rho}\sin(\Delta m
\Delta t) \;\; , \label{10}
\end{equation}
where
\begin{equation}
C = \frac{1-|\lambda|^2}{1+|\lambda|^2} \; , \;\; S = \frac{2Im
\lambda}{1+|\lambda|^2} \; , \;\; \lambda = \frac{q}{p} \frac{\bar
A_{\rho^+ \rho^-}}{A_{\rho^+ \rho^-}} \;\; , \label{11}
\end{equation}
factor $\frac{q}{p} = e^{-2i\beta}$ appears from $B_d - \bar B_d$
mixing. Here are the experimental data \cite{6}:
\begin{eqnarray}
C_{\rho^+ \rho^-} & = & -0.23 \pm 0.24 \pm 0.14 \;\; , \nonumber
\\
S_{\rho^+ \rho^-} & = & -0.19 \pm 0.33 \pm 0.11 \;\; . \label{12}
\end{eqnarray}

The smallness of the penguin contribution is manifested in the
smallness of the $C_{\rho^+ \rho^-}$ value in comparison with 1
and we see that even the value $C_{\rho^+ \rho^-} =0$ (and
$P/T=0$) does not contradict the data. Neglecting penguin
amplitude
\begin{equation}
S_{\rho^+ \rho^-} = \sin 2\alpha = -0.19 \pm 0.35 \; , \;\; \alpha
= 95^o \pm 10^o \;\; , \label{13}
\end{equation}
where theoretical systematic uncertainty due to nonzero $P/T$
ratio is omitted.

\section{$B \to \pi\pi$}

As it was demonstrated in paper \cite{7} from the experimental
data on averaged branching ratios and asymmetries of $B_d(\bar
B_d) \to \pi^+ \pi^-$, $\pi^0 \pi^0$ and $B_u \to \pi^+ \pi^0$
decays one can extract angle $\alpha$ relying only on isospin
relations for decay amplitudes. However, as it was noticed in the
same paper, one should expect large experimental uncertainties in
the parameters describing the decays to the pair of neutral pions
which will prevent direct determination of $\alpha$ with good
accuracy. And this really happens. Unfortunately unlike the case
of $\rho\rho$ decays, the branching ratio of $B_d(\bar B_d) \to
\pi^0 \pi^0$ is comparable to that to charged modes preventing
bounding the penguin contributions to $B\to\pi\pi$ decays. Let us
note that the data of Belle and BaBar on $\pi^+ \pi^-$  and
$\pi^\pm \pi^0$ branching ratios well agree, while their
difference in $\pi^0 \pi^0$ branching ratio is within two standard
deviations. However, considerable $B_d(\bar B_d) \to \pi^0\pi^0$
branching ratio did not necessary mean that the penguin
contribution is comparable to a tree one. In order to investigate
how large it is let us look at experimental data on $C_{\pi^+
\pi^-}$ and $S_{\pi^+ \pi^-}$. Here the data of Belle and BaBar
are in disagreement \cite{8}:
\begin{equation}
\begin{array}{lcc}
 & {\rm BaBar} & {\rm Belle} \\
C_{\pi^+ \pi^-} & -0.09 \pm 0.16 & -0.56 \pm 0.14 \\ S_{\pi^+
\pi^-} & -0.30 \pm 0.17 & -0.67 \pm 0.17
\end{array}
\label{14}
\end{equation}
According to BaBar data the tree amplitude dominates in the decay
to $\pi^+\pi^-$ ($C_{\pi^+\pi^-} \approx 0$)
while according to Belle this is not so
($C_{\pi^+\pi^-}$ differs from zero by four sigmas).
That is why analyzing data two strategies were implied: the Belle
and BaBar data were either averaged (see for example \cite{9}) or
disregarded.

We do not want to average data which contradict each other,
neither we want to disregard them. Instead we suppose that BaBar
data are correct, not Belle. As an argument in favor of this
statement we can suggest the results of paper \cite{10}, where the
contributions of QCD penguin diagrams to $B\to \rho\rho, \rho\pi,
\pi\pi$ decays were found to be small. Neglecting them, from BaBar
measurement of $S_{\pi^+ \pi^-}$ we obtain:
\begin{equation}
\sin 2\alpha = S_{\pi^+ \pi^-} = -0.30 \pm 0.17 \; , \;\; \alpha =
99 \pm 5^o \;\; . \label{15}
\end{equation}

\section{$B \to \rho \pi$}

The time dependence of these decays is given by the following
formula \cite{15'}:
\begin{eqnarray}
\frac{dN(B_d(\bar B_d)\to\rho^\pm \pi^\mp)}{d \Delta t} & = &(1\pm
A_{CP}^{\rho\pi})e^{-\Delta t/\tau}[1-q(C_{\rho\pi} \pm \Delta
C_{\rho\pi}) \cos(\Delta m \Delta t) + \nonumber \\ & + &
q(S_{\rho\pi} \pm \Delta S_{\rho\pi})\sin(\Delta m \Delta t)] \;\;
, \label{16}
\end{eqnarray}
where $q = -1$ describes the $B_d$ decay probability dependence on
$\Delta t$ (at $\Delta t =0$ $\bar B_d$ decays) and $q=1$
corresponds to $\bar B_d$ decay probability dependence on $\Delta
t$ (at $\Delta t =0$ $B_d$ decays); $$A_{CP}^{\rho\pi} =
\frac{|A^{+-}|^2 - |\bar A^{-+}|^2 + |\bar A^{+-}|^2 -
|A^{-+}|^2}{|A^{+-}|^2 + |\bar A^{-+}|^2 + |\bar A^{+-}|^2 +
|A^{-+}|^2} \;\; , $$
\begin{equation}
C_{\rho\pi}\pm \Delta C_{\rho\pi} = \frac{|A^{\pm\mp}|^2 - |\bar
A^{\pm\mp}|^2}{|A^{\pm\mp}|^2 + |\bar A^{\pm\mp}|^2} \;\;
\label{16'}
\end{equation}
$$ S_{\rho\pi} \pm \Delta S_{\rho\pi} = \frac{2 {\rm
Im}(\frac{q}{p} \frac{\bar
A^{\pm\mp}}{A^{\pm\mp}})}{1+\left|\frac{\bar
A^{\pm\mp}}{A^{\pm\mp}}\right|^2} \;\; . $$

The amplitudes $A^{\pm\mp}$ describe $B_d$ decays, $\bar
A^{\pm\mp} - \bar B_d$ decays and the first sign is that of
produced $\rho$-meson (for example $A^{+-}$ is the amplitude of
$B_d \to \rho^+ \pi^-$ decay). It is convenient to write the
amplitudes of $B\to \rho\pi$ decays as sums of tree and penguin
contributions:
\begin{equation}
\begin{array}{ll}
A^{+-} = A_1 e^{i\delta_1} e^{i\gamma} + P_1 e^{i\delta_{P_1}} \;
, & A^{-+} = A_2 e^{i\delta_2} e^{i\gamma} + P_2 e^{i\delta_{P_2}}
\;\; , \\ \bar A^{-+} = A_1 e^{i\delta_1} e^{-i\gamma} + P_1
e^{i\delta_{P_1}} \; , & \bar A^{+-} = A_2 e^{i\delta_2}
e^{-i\gamma} + P_2 e^{i\delta_{P_2}} \;\; , \\
\end{array}
\label{16''}
\end{equation}
where amplitude $A_1$ corresponds to $\rho$-meson produced from
$W$-boson ($b\to u\rho^-$, $\bar b \to \bar u \rho^+$) and
amplitude $A_2$ describes $\pi$-meson produced from $W$-boson
($b\to u\pi^-$, $\bar b \to \bar u\pi^+$). Amplitude $P_1$
corresponds to a penguin diagram in which a spectator quark is
involved in $\pi$-meson production, while $P_2$ -- to
participation of a spectator quark in $\rho$-meson production.

If one can neglect penguin amplitudes, then formulas for physical
observables are as follows: $$ A_{CP}^{\rho\pi} = C_{\rho\pi} = 0
\; , \;\; \Delta C_{\rho\pi} = \frac{A_1^2 - A_2^2}{A_1^2 + A_2^2}
\; , \;\; \Delta S_{\rho\pi} = \frac{2A_1 A_2}{A_1^2 + A_2^2}
\sin(\delta_2 - \delta_1)\cos 2\alpha $$
\begin{equation}
S_{\rho\pi} = \frac{2A_1 A_2}{A_1^2 + A_2^2}\cos(\delta_2
-\delta_1)\sin 2\alpha \;\; . \label{17}
\end{equation}
We use the results of fits of the $\Delta t$ distributions
obtained by Belle \cite{11} and BaBar \cite{12}.

Let us start from the experimental measurements of parameter
$C_{\rho\pi}$:
\begin{equation}
C_{\rho\pi} = \begin{array}{ll} 0.25 \pm 0.17 \; , & {\rm Belle}
\;\; , \\ 0.34 \pm 0.12 \; , & {\rm BaBar} \;\; , \end{array}
\label{18}
\end{equation}
and we see that while Belle result is compatible with the
hypothesis that $P/T \ll 1$, BaBar result almost contradicts it.
Waiting for more precise data let us go on supposing that penguin
contribution is negligible. By the way, the data on $A_{CP}$
confirms the smallness of penguin amplitude:
\begin{equation}
A_{CP}^{\rho\pi} = \begin{array}{ll} -0.16 \pm 0.10 \; , & {\rm
Belle} \;\; , \\ -0.088 \pm 0.051 \; , & {\rm BaBar} \;\; .
\end{array}
\label{19}
\end{equation}

For $\Delta C_{\rho\pi}$ the result is:
\begin{equation}
\Delta C_{\rho\pi} = \begin{array}{ll} 0.38 \pm 0.18 \; , & {\rm
Belle} \;\; , \\ 0.15 \pm 0.12 \; , & {\rm BaBar} \;\; ,
\end{array}
(\Delta C_{\rho\pi})_{\rm average} = 0.22 \pm 0.10 \label{20}
\end{equation}
and it means that $A_1 \approx 1.3 A_2$.

For $\Delta S_{\rho\pi}$ we have:
\begin{equation}
\Delta S_{\rho\pi} = \begin{array}{ll} -0.30 \pm 0.25 \; , & {\rm
Belle} \;\; , \\ 0.22 \pm 0.15 \; , & {\rm BaBar} \;\; ,
\end{array}
(\Delta S_{\rho\pi})_{\rm average} = 0.08 \pm 0.13 \label{21}
\end{equation}
and its smallness means that $\sin(\delta_1 - \delta_2) \approx 0$
(another solution, $\cos 2\alpha \approx 0$, is unacceptable). It
means that both $\delta_1$ and $\delta_2$ are small, or that they
are close to each other. Substituting  $\cos(\delta_1
-\delta_2)=1$ into expression for $S_{\rho\pi}$ we obtain:
\begin{equation}
S_{\rho\pi} = \sqrt{1-(\Delta C_{\rho\pi})^2}\sin 2\alpha \;\; ,
\label{22}
\end{equation}
while the experimental results are:
\begin{equation}
S_{\rho\pi} = \begin{array}{ll} -0.28 \pm 0.25 \; , & {\rm Belle}
\;\; , \\ -0.10 \pm 0.15 \; , & {\rm BaBar} \;\; ,
\end{array} (S_{\rho\pi})_{\rm average} = -0.15 \pm 0.13
\label{23}
\end{equation}

From (\ref{20}), (\ref{22}) and (\ref{23}) we obtain:
\begin{equation}
\alpha = 94^o \pm 4^o \;\; . \label{24}
\end{equation}

\section{Conclusion}

Averaging the results for $\alpha$ presented in eqs.(\ref{13}),
(\ref{15}) and (\ref{24}) we obtain:
\begin{equation}
\alpha = 96^o \pm 3^o \;\; , \label{25}
\end{equation}
where only the experimental error is taken into account, while the
theoretical uncertainty coming from the penguin diagrams is
neglected. Let us note that result (\ref{25}) is in good agreement
with global CKM fit results:
\begin{equation}
\alpha_{\rm UTfit}^{\rm \cite{13}} = 94^o \pm 8^o \; , \alpha_{\rm
CKMfitter}^{\rm \cite{14}} = 94^o \pm 10^o \;\; , \label{26}
\end{equation}
$$\alpha_{\rm AOV}^{\rm \cite{15}} = 100^o \pm 5^o \;\; . $$

How large can penguin contributions be in comparison with tree
ones?

Strong interaction renormalization for beauty hadron weak decays
is much smaller than for strange particles, because the masses of
beauty hadrons are much closer to $M_W$ in the logarithmic scale.
Here are the results of NLO calculations from Table 1 of paper
\cite{18}, where we take numbers which correspond to the modern
value of $\alpha_s(M_Z) = 0.12$ ($\Lambda_4 = 280$ MeV): $$ c_2 =
1.14 \; , \;\; c_1 = -0.31 \; , \;\; c_3 = 0.016 \; , \;\; c_5 =
0.010 \; , c_4 = -0.036 \; , $$
\begin{equation}
c_6 = -0.045 \label{28}
\end{equation}
and we observe that the renormalization coefficients of penguin
operators ($O_3 - O_6$) do not exceed 4\% of that for tree-level
operator ($O_2$). Concerning the matrix elements one can
definitely state that a large enhancement factor $m_\pi^2/(m_u +
m_d) m_s \approx 10$ which makes penguins so important in
explaining $\Delta I = 1/2$ rule in nonleptonic weak decays of
strange particles is absent in beauty hadron decays, being
substituted by $m_\pi^2/(m_u + m_d)m_b \approx 1/2$.

A grain of salt comes from the CKM matrix elements which enhance
the penguin amplitude with respect to the tree one by factor
$(\rho^2 + \eta^2)^{-0.5} \approx 2$\footnote{We are grateful to
P.N. Pakhlov for this remark.}.

It follows that the theoretical uncertainty in (\ref{25}) coming
from the penguin diagrams can be close to the experimental one.

\section*{Acknowledgements}
M.V. is grateful to A.B. Kaidalov for useful discussions and to
CERN TH, where a part of the work was done, for hospitality. This
work was partially supported by the program FS NTP FYaF
40.052.1.1.1112 and by grant NSh-2328.2003.2. G.O. is grateful to
Dynasty Foundation for partial support.


\begin{thebibliography}{99}
\bibitem{1} K. Abe et al. (Belle Collaboration), hep-ex/0408111;
\\
B. Aubert et al. (BaBar Collaboration), hep-ex/0408127 (2004).

\bibitem{2} Y. Grossman and H.R. Quinn, Phys. Rev. {\bf D58}
(1998) 017504.

\bibitem{3} B. Aubert et al. (Babar Collaboration), Phys. Rev.
Lett. {\bf 93} (2004) 231801.

\bibitem{4} B. Aubert et al. (Babar Collaboration), Phys. Rev.
Lett. {\bf 91} (2003) 171802.

\bibitem{5} K. Abe et al. (Belle Collaboration), Phys. Rev. Lett.
{\bf 91} (2003) 221801.

\bibitem{6} B. Aubert et al. (Babar Collaboration), hep-ex/0407051
(2004).

\bibitem{13} J. Charles et al. (The CKMfitter Group),
hep-ph/0406184 (2004).

\bibitem{14} M. Bona et al. (UTfit Collaboration),
hep-ph/0408079 (2004).

\bibitem{15} E.A. Andriyash, G.G. Ovanesyan and M.I. Vysotsky,
hep-ph/0502111 (2005).

\bibitem{7} M. Gronau and D. London, Phys. Rev. Lett. {\bf 65}
(1990) 3381.

\bibitem{8} B. Aubert et al. (Babar Collaboration), hep-ex/0501071
(2005); \\ K. Abe et al. (Belle Collaboration), hep-ex/0502035 (2005).

\bibitem{9} A. Ali, E. Lunghi and A.Ya. Parkhomenko, Eur. Phys. J.
{\bf C36} (2004) 183; \\ A.Ya. Parkhomenko, hep-ph/0411061 (2004);
\\ M. Gronau, hep-ph/0407316 (2004).

\bibitem{10} R. Aleksan et al., Phys. Lett. {\bf B356} (1995) 95.

\bibitem{15'} M. Gronau, Phys. Lett. {\bf B233} (1989) 479.

\bibitem{11} C.C. Wang et al. (Belle Collaboration),
hep-ex/0408003 (2004).

\bibitem{12} B. Aubert et al. (Babar Collaboration),
hep-ex/0408099 (2004).

\bibitem{18} A.J. Buras, M.~Jamin, M.E.~Lautenbacher and
P.H.~Weisz, Nucl. Phys. {\bf B370} (1992) 69.
\end{thebibliography}
\end{document}